\documentclass[%
 reprint,
 amsmath,amssymb,
 aps,
]{revtex4-2}

\usepackage{graphicx}
\usepackage{dcolumn}
\usepackage{bm}

\usepackage{color}

\begin{document}


\title{Enzymatic Mpemba Effect:\\ Slowing of biochemical reactions by increasing enzyme concentration}

\author{Tetsuhiro S. Hatakeyama}
 \email{hatakeyama@elsi.jp}
 \affiliation{Earth-Life Science Institute (ELSI), Institute of Science Tokyo, 2-12-1-IE-1 Ookayama, Meguro-ku,
Tokyo 152-8550, Japan}

\date{\today}

\begin{abstract}
Increasing the enzyme concentration generally speeds up enzymatic reactions. 
However, in this Letter, we show that increasing the enzyme concentration can also slow down the enzymatic reaction.
We consider a simple allosteric protein with multiple modification sites, catalyzed by two enzymes with the same catalytic activity, but slightly different affinities.
We show that increasing the concentration of one enzyme can slow the relaxation to the equilibrium state. 
The mechanism for this slowing is similar the Markovian Mpemba effect, and we name this phenomenon as the Enzymatic Mpemba effect.
\end{abstract}

\maketitle


Most biochemical reactions require enzymes, because, the activation energy for the typical biochemical reaction is too high to progress in observable timescales without a catalyst.
In general, a reaction kinetics of biochemical reactions with enzymes is described by the Michaelis-Menten kinetics \cite{Michaelis1913} given by
\begin{equation}
v = \frac{k [E] [S]}{K + [S]},
\end{equation}
where $k$ is a kinetic rate of the elemental reaction, $[E]$ and $[S]$ are concentrations of an enzyme and substrate, respectively, and $K$ is a dissociation constant between the enzyme and substrate, which is microscopically originated by a binding energy between the enzyme and substrate \cite{Phillips2012}.
From the Michaelis-Menten equation, when the concentration of the substrate is much lower than the dissociation constant, the reaction rate is approximately given by $v \simeq k [E] [S]/ K$.
In contrast, when the concentration of the substrate is much higher than the dissociation constant, the reaction rate is approximately given by $v \simeq k [E]$, i.e., the reaction speed is limited by the enzyme binding.
In any cases, an increase in the enzyme concentration accelerates the enzymatic reactions.

In this Letter, we show that increasing the enzyme concentration can slow down the enzymatic reaction. 
We consider a biomolecule with multisite modifications catalyzed by enzymes.
If there is only one type of enzyme, i.e. one enzyme catalyzes all the modification reactions, the relaxation time to steady state is inversely proportional to the enzyme concentration.
This is consistent with the conventional picture given by the Michaelis-Menten equation.
In contrast, if two enzymes are present, an increase in the concentration of one enzyme can increase the relaxation time to steady state, even if both enzymes have the same catalytic activity.

We found that this anomalous deceleration is analogous to the Markovian Mpemba effect \cite{Lu2017}, a class of Mpemba effects in which matter initially hot can become cold faster than matter initially colder \cite{Mpemba1969}. 
In the relaxation process from an equilibrium state at one temperature to that at a lower temperature, matter can, contrary to intuition, relax faster when the initial temperature is higher. 
This is the phenomenon addressed in the Markovian Mpemba effect. 
This counter-intuitive phenomenon is explained by the distance to the final equilibrium state from the initial equilibrium state at the higher temperature is closer than that from the initial state at the lower temperature \cite{Lu2017}.
The presence of an enzyme does not change the equilibrium state because the enzyme is catalyst, and a mechanism of slowing down of biochemical reactions is not exactly the same as the Markovian Mpemba effect. 
Instead of the equilibrium state, the enzymatic system stays in a plateau for a long time before relaxing to the equilibrium state.
The distance to the equilibrium state from the plateau with the lower enzyme concentration is closer than that from the plateau with the higher enzyme concentration.
Therefore, based on this analogy, we term the slowing down in the enzymatic system depending on the increase in the enzyme concentration as the \textit{Enzymatic Mpemba effect}.


\begin{figure}[tbhp]
\centering
\includegraphics[width=1.0\linewidth]{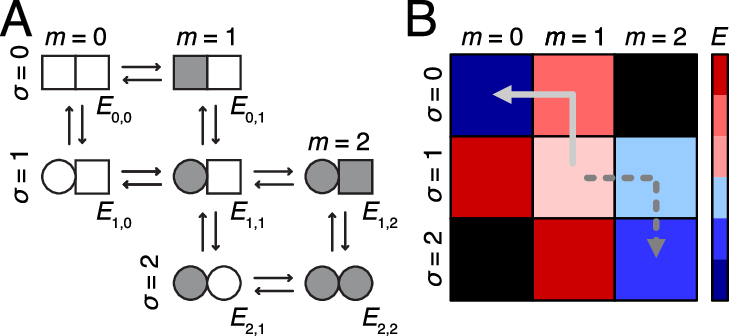}
\caption{
Schematic representation of an allosteric model for a dimeric protein.
(A) Color of each monomer represents a modification state: white and filled ones are unmodified and modified monomer, respectively.
Each square or circle represents a monomer with different structure: Square and circle ones are tence and relaxed monomer, respectively.
Transitions along horizontal arrows and vertical arrows are modification reactions and structural changes, respectively.
(B) Heatmap of the energy of each state, $E_{i,j}$. 
Color of each state corresponds to the energy indicating by the color bar.
The blacked-out state is unavailable.
Light-gray solid and dark-gray dashed lines indicate the stable and metastable paths, respectively (see the main text).
\label{fig:model}
}
\end{figure}

We consider an allosteric protein. 
Models for the allosteric proteins are divided into some classes, a concerted model (MWC model: Monod-Wyman-Changeux model) \cite{Monod1965}, a sequential model (KNF model: Koshland-N\'{e}methy-Filmer model) \cite{Koshland1966}, or a mixture of these \cite{Hilser2012}.
Here, we employ the sequential model (Fig. \ref{fig:model}).
In this model, a protein consists of $M$ monomers, which have a modification site and two different structures.
When a monomer is modified (color change in Fig. \ref{fig:model}A), it tends to change its structure from one to the other (shape change in Fig. \ref{fig:model}A); We call these two states as tense and relaxed state.
We represent the number of relaxed and modified monomers in a single molecule as $\sigma$ and $m$, respectively.
In the original KNF model, the modification and structural change are assumed to occur simultaneously, i.e., when a monomer is modified, it immediately transitions between the tense and relaxed states.
Thus, $\sigma$ and $m$ should be identical.
However, those two processes should be distinguished, because the former is enzyme-independent while the latter is enzyme-dependent.
Hence, $\sigma$ and $m$ will take different values in actual molecules, and we allow $\sigma$ to take $\sigma = m \pm 1$ by following the original idea of the KFN model.
For the sake of simplicity, we here consider a dimeric protein.
Thus, we should consider seven states of the protein and transitions among those states.
We give the energy of each state of the molecule as $E_{i, j}$ for a molecule in the $\sigma = i, m = j$ state.
We set $E_{0, 0} = 0$, $E_{0, 1} = 4$, $E_{1, 0} = 5$, $E_{1, 1} = 3$, $E_{1, 2} = 2$, $E_{2, 1} = 5$, and $E_{2, 2} = 1$ (Fig. \ref{fig:model}B).

In general, the activation energy for biochemical reactions is too high to occur the reactions in the observation time without any catalyst, while the enzyme can drastically decrease the activation energy.
We assumed that when an enzyme does not bind, modification reactions do not progress.
We consider a case where two enzymes catalyze the same reaction; we refer these two enzymes as enzyme 1 and 2 for convenience.
For the sake of simplicity, we assume that those have the same activity but only the affinity to the substrate is different, which depends on the structure but not on the modification state.
We describe a binding energy of the enzyme $j$ to a substrate in the $\sigma = i$ state as $E^{\mathrm{b}j}_i$ and set them as $E^{\mathrm{b}1}_0 = 0.0$, $E^{\mathrm{b}1}_1 = 1.5$, $E^{\mathrm{b}1}_2 = 5.0$ for the enzyme 1, and $E^{\mathrm{b}2}_0 = 1.5$, $E^{\mathrm{b}2}_1 = 0.0$, $E^{\mathrm{b}2}_2 = 5.0$ for the enzyme 2.
That is, two enzymes have slightly different substrate preferences, but it should be noted that both enzymes can catalyze all reactions independently.
Thus, the transition probability for modification reactions is derived:
\begin{widetext}
\begin{equation}
\begin{split}
P \left(\sigma = i, m = i \pm 1 | \sigma = i, m = i \right) &= \left[ 1 - \left(1 - p^\mathrm{b1}_i \right) \left(1 - p^\mathrm{b2}_i \right) \right]  \times \min \left\{1, \exp\left[ -\beta \left(E_{i, i \pm 1} - E_{i, i} \right) \right] \right\}, \\
P \left(\sigma = i, m = i | \sigma = i, m = i \pm 1 \right) &= \left[ 1 - \left(1 - p^\mathrm{b1}_i \right) \left(1 - p^\mathrm{b2}_i \right) \right] \times \min \left\{ 1, \exp\left[ -\beta \left(E_{i, i} - E_{i, i \pm 1} \right) \right] \right\},
\end{split}
\end{equation}
\end{widetext}
We set the Boltzmann constant as an unity, and thus $\beta$ is inverse temperature and fixed as 1.5.
$p^\mathrm{b1}_i$ and $p^\mathrm{b2}_i$ are binding probabilities of the enzyme 1 and 2 to a substrate in the $\sigma = i$ state, respectively, and are derived below (see the supplementary information for the detailed derivation) :
\begin{equation}
p^{\mathrm{b}j}_i = \frac{\exp(\beta \mu)}{\exp\left( -\beta E^{\mathrm{b}j}_i \right) + \exp(\beta \mu)}.
\end{equation}

In contrast, the structural change does not require the enzyme.
Thus, the transition probability is given as
\begin{widetext}
\begin{equation}
\begin{split}
&P \left(\sigma = i \pm 1, m = i | \sigma = i, m = i \right) = \min \left\{ 1, \exp\left[ -\beta \left(E_{i \pm 1, i} - E_{i, i} \right) \right] \right\}, \\
&P \left(\sigma = i, m = i | \sigma = i \pm 1, m = i \right) = \min \left\{ 1, \exp\left[ -\beta \left(E_{i, i} - E_{i \pm 1, i} \right) \right] \right\}.
\end{split}
\end{equation}
\end{widetext}
Note that the binding probabilities $p^\mathrm{b1}_i$ and $p^\mathrm{b2}_i$ are increasing functions of the number of enzymes, $N_\mathrm{Enz1}$ and $N_\mathrm{Enz2}$ for the enzyme 1 and 2, respectively.
Hence, if the number of enzymes increases, the speed of modification reactions should be fastened.

Both modification and structural changes satisfy the detailed balance condition because the enzyme does not alter the equilibrium.
Hence, the number of molecules in the $\sigma = i, m = j$ state, $N_{i, j}$, in the equilibrium, $N^{\rm eq}_{i,j}$, is proportional to $\mathcal{N}_{i,j} \exp(-\beta E_{i,j})$.
$\mathcal{N}_{i,j}$ is the number of the combination of molecules with the same $\sigma$ and $m$, represented by $\mathcal{N}_{i,j} = \max \left[ \tbinom{M}{i}, \tbinom{M}{j} \right]$.

We observed the relaxation process from the initial state to the equilibrium state. 
We calculated the relaxation dynamics by the Monte Carlo method, which give the same time course of the corresponding model described by ordinary differential equations on average, as previously reported (see \cite{Hatakeyama2020}).
We define the timescale of each reaction as 1 second, and thus 1 Monte Carlo step (MCS) corresponds to 1 second.
As an initial condition, all molecules are in the $\sigma = 1, m = 1$ state. 
We set the total number of molecules, $N = \Sigma_{i, j} N_{i,j}$, as 100.


\begin{figure}[tbhp]
\centering
\includegraphics[width=1.0\linewidth]{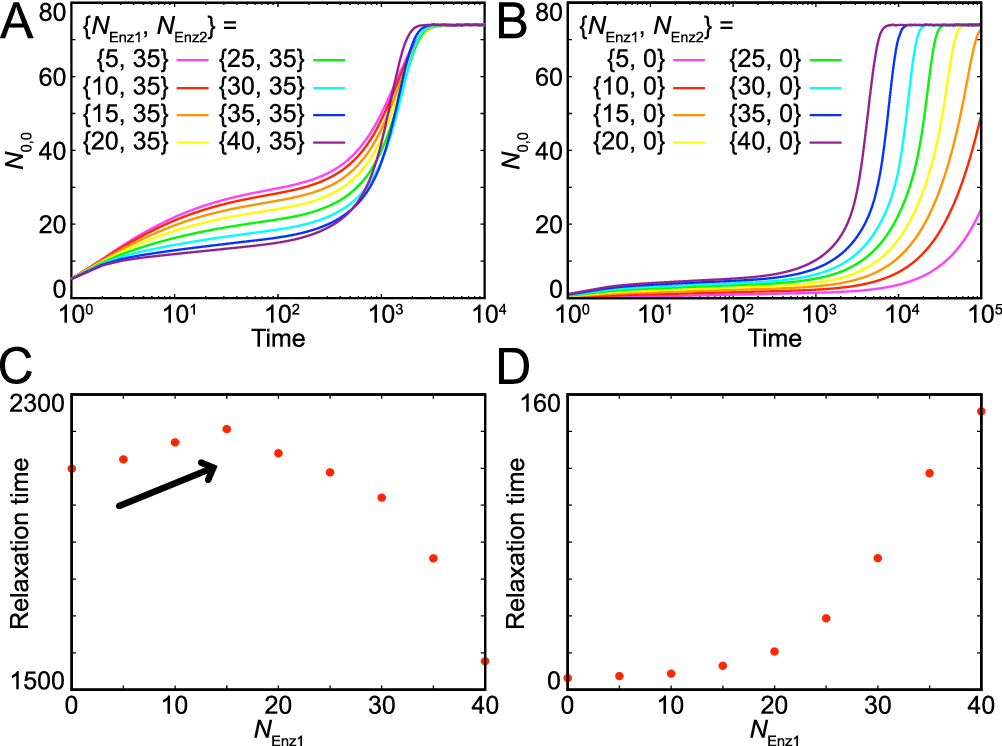}
\caption{
Increase in $N_{\rm Enz1}$ slows down the relaxation. 
(A, B) Relaxation process of $N_{0,0}$ with varied $N_{\rm Enz1}$ and fixed $N_{\rm Enz2}$ as (A) 35 and (B) 0. 
Different colored line indicates the time evolution with different $N_{\rm Enz1}$. 
Each line is average over 1000 realizations. 
(C, D) Relaxation time in which (C) $N_{0,0}$ reached its equilibrium value $N_{0,0}$ and (D) 15. 
Each point is an averaged value over 1000 realization. 
\label{fig:relaxation}
}
\end{figure}

Although the distribution of molecules in the equilibrium is independent of the number of enzymes, the time course of the number of molecule depends on the number of enzymes.
The number of the $\sigma = 0, m = 0$ state molecule, $N_{0,0}$, which is the most abundant component in the equilibrium, showed the multi step relaxation when catalyzed by both enzymes 1 and 2 (Fig. \ref{fig:relaxation}A).
$N_{0,0}$ increased at first, then the rate of increase slowed and $N_{0,0}$ showed a plateau.
Then, after the plateau, $N_{0,0}$ relaxes to the equilibrium value $N^{\rm eq}_{0,0}$.

When both enzymes are present, increasing $N_{\rm Enz1}$ causes a slower relaxation to equilibrium in a region with small $N_{\rm Enz1}$ (Fig. \ref{fig:relaxation}C), even though the enzyme 1 can catalyze all reactions. 
It was more apparent in the initial increase in $N_{0,0}$;
The time for $N_{0,0}$ to reach 15 was slower as $N_{\rm Enz1}$ increased, even though $N_{\rm Enz1}$ increases more than $N_{\rm Enz2}$ (Fig. \ref{fig:relaxation}D).
This indicates that there is a counter-intuitive phenomenon in which the increase in enzymes slows down the relaxation of the system. 

Note that this slowing down was observed when two enzymes existed.
If only the enzyme 1 was present and not the enzyme 2, an increase in $N_{\rm Enz1}$ led to faster relaxation ordinarily (Fig. \ref{fig:relaxation}B).

\begin{figure}[tbhp]
\centering
\includegraphics[width=1.0\linewidth]{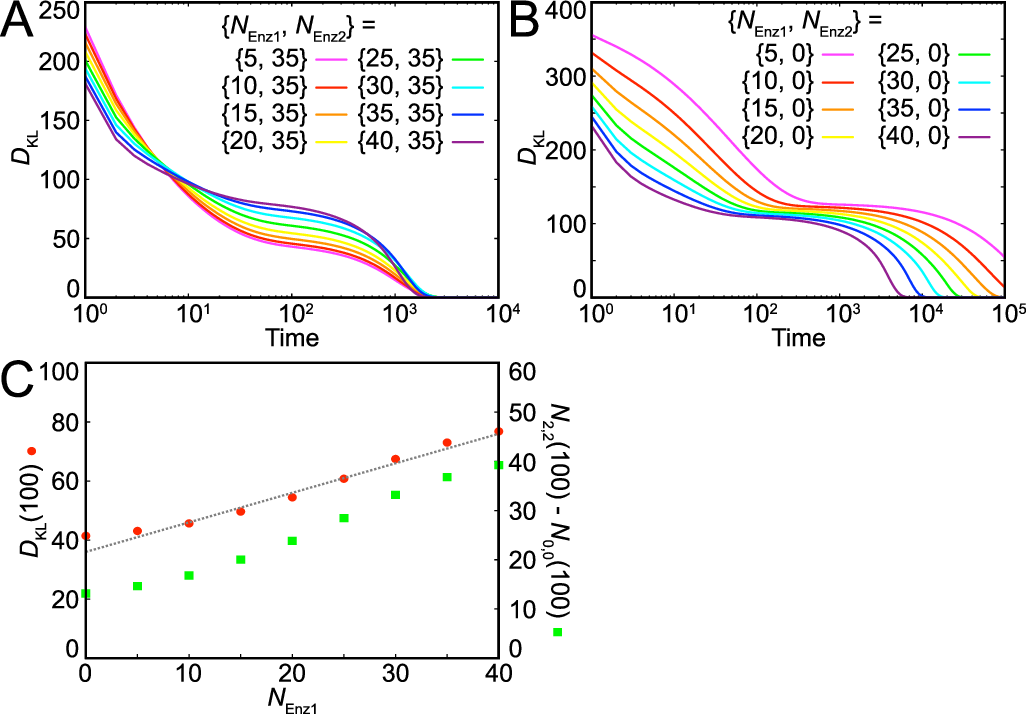}
\caption{
Increase in $N_{\rm Enz1}$ shifts the metastable state away from the equilibrium state.
(A, B) Time evolution of $D_{\rm KL}$ with varied $N_{\rm Enz1}$ and fixed $N_{\rm Enz2}$ as (A) 35 and (B) 0. 
(C) $D_{\rm KL}$ (red circles) and the difference between $N_{2,2}$ and $N_{0,0}$ (green squares) at $t = 100$ with varied $N_{\rm Enz1}$. 
$N_{\rm Enz2}$ is fixed as 35 as the same as (A). 
Gray dotted line indicates $N_{\rm Enz1} + 35$ for $D_{\rm KL}$.
\label{fig:kullback_leibler}
}
\end{figure}

To investigate why the slowing down occurred, we introduce the Kullback-Leibler divergence between distributions of modification states of molecules during the relaxation to that in the equilibrium as
\begin{equation}
D_{\rm KL}(t) = \sum_{i+j+k=N} p_{i, j, k}(t) \frac{p_{i,j,k}(t)}{p^{\rm eq}_{i,j,k}},
\end{equation}
where $p_{i,j,k}(t)$ is the probability at time $t$ that the number of molecules in $m = 0$, $1$, and $2$ states is $i$, $j$, and $k$, respectively, and $p^{\rm eq}_{i,j,k}$ is the probability at the equilibrium.
Since all of reactions satisfy the detailed balance condition, the above $D_{\rm KL}$ is positive, decreases monotonically with time, and finally becomes zero at the equilibrium state, i.e., it works as the Lyapnov function \cite{vanKampen1959, vanKampen1992}.
Then, $D_{\rm KL}$ is an indicator of the distance to equilibrium.

The relaxation of the KL divergence also showed the plateau (Fig. \ref{fig:kullback_leibler}).
When both enzymes are present, $D_{\rm KL}$ at the plateau increased with $N_{\rm Enz1}$ (Fig. \ref{fig:kullback_leibler}A).
Indeed, $D_{\rm KL}$ at the plateau increased with $N_{\rm Enz1}$ almost proportional to that (red circles in Fig. \ref{fig:kullback_leibler}C).
This increase in $D_{\rm KL}$ at the plateau correlated with $N_{2,2} - N_{0,0}$ (green squares in Fig. \ref{fig:kullback_leibler}C).
It implies that the increase in the distance between the plateau and the equilibrium state is due to a transient increase in $N_{2,2}$.
Note that when only the enzyme 1 is present, although the plateau itself was present, an increase in $N_{\rm Enz1}$ brought the plateau closer to equilibrium (Fig. \ref{fig:kullback_leibler}B) and monotonically decreased the relaxation time.
Therefore, two key factors in the slowing down are the existence of the plateau and the increase in the distance between the plateau and the equilibrium due to the increase in the enzyme amount.

Why does the system stay in the plateau with a large $N_{2,2}$?
Since $E_{0,0} < E_{2,2} < E_{1,1}$ is satisfied, molecular states show a multistable landscape at the single molecular level, and the $\sigma = 2, m = 2$ state is metastable (Fig. \ref{fig:model}B). 
Here, the affinity of enzymes for the $\sigma = 2$ state molecule is much higher than for other states of molecules, and when $N_{2, 1} + N_{2, 2}$ is larger than the number of enzymes, enzymes are sequestrated by these molecules, and $p^{\mathrm{b}i}_0$ and $p^{\mathrm{b}i}_1$ for both enzymes get smaller.
Then, reactions from the $\sigma = 1$ to $\sigma = 0$ states are drastically slowed down, and the system stays in the plateau for a long time.
We term this state as a kinetic-memory state because this state memorizes the initial concentration of enzymes. 
The kinetic-memory state exists as a continuous region in the phase space in the limit of $N \rightarrow \infty$: 
If $N_{2,2}$ exceeds the number of enzymes, any molecular distribution in the $\sigma = 0$ and $\sigma = 1$ states is kept as a kinetic-memory state. 
Even if the number of molecules is finite, a state with sufficiently large $N_{2,2}$ will exist as the kinetic-memory state.
Such a mechanism is similar to the previously reported enzymatic kinetically constrained model (eKCM) \cite{Hatakeyama2020, Hatakeyama2014}.

The relaxation to the equilibrium should occur via the kinetic-memory state.
In the early relaxation process, each $\sigma = 1, m = 1$ state molecule relaxes to the $\sigma = 0, m = 0$ or $\sigma = 2, m = 2$ states.
Here, $E_{1,0}$ and $E_{2,1}$ are higher than $E_{0,1}$ and $E_{1,2}$, and then molecules tend to relax to the $\sigma = 0, m = 0$ state via the $\sigma = 0, m = 1$ state (stable path; light-gray solid line in Fig. \ref{fig:model}B) or the $\sigma = 2, m = 2$ state via the $\sigma = 1, m = 2$ state (metastable path; dark-gray dashed line in Fig. \ref{fig:model}B).
Since the energy of the former intermediate state ($E_{1,2}$) is lower than that for the latter state ($E_{0,1}$), the molecule tends to be the $\sigma = 2, m = 2$ state more than the $\sigma = 0, m = 0$ state.

Finally, why does increasing in the number of enzyme push the kinetic-memory state away from equilibrium?
The two enzymes have slightly different affinities for different molecule states: Enzyme 1 shows higher affinity for a $\sigma = 1$ state  molecule than for a $\sigma = 0$ state molecule, and the enzyme 2 shows the inverse affinity, i.e., enzymes 1 and 2 tend to catalyze the modification reactions of the $\sigma = 1$ and $\sigma = 0$ state molecules, respectively.
Since, the stable and metastable paths include the modification reactions of the $\sigma = 0$ and $\sigma = 1$ molecules, respectively, an increase in $N_{\rm Enz1}$ with the fixed $N_{\rm Enz2}$ changes the balance of those paths to increase the flux of the metastable path and make the kinetic-memory state far from the equilibrium.
Therefore, the increase in the enzyme can slow down the relaxation even though it speeds up each reaction.
Note that if there is only one enzyme, an increase in it does not change the balance in branched reactions and never slows down the relaxation.


In this Letter, we theoretically showed the existence of a counter-intuitive phenomenon: increasing the enzyme concentration slows down the enzymatic reaction.
From the mechanism discussed above, below conditions are required: \\
1) Modification reactions catalyzed by enzymes show a branched network structure, and one of the branches provides the local energy minimum. \\
2) Substrates in the local minimum sequestrate enzymes, and a kinetic-memory state exists. \\
3) Two (or more) enzymes have different affinities for substrates, and one enzyme binds preferentially to a substrate that exhibits an enzymatic reaction toward the local minimum. \\
We have shown that the minimal model in which these conditions are satisfied is an allosteric dimeric protein. 
Of course, the probability of satisfying these conditions will increase in more complex proteins. 
In particular, for proteins related in the cell signaling, some enzymes catalyzes multiple modification sites on a substrate \cite{Yang2005, Salazar2009, Mylona2016}, and the phenomenon we have discovered may be actively employ to control the timescale of signal transduction.

\begin{figure}[tbhp]
\centering
\includegraphics[width=0.9\linewidth]{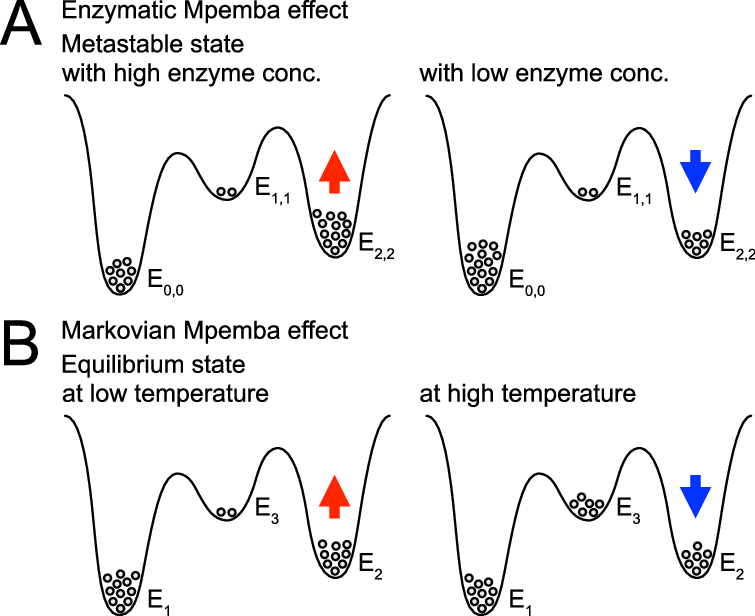}
\caption{
Comparison of (A) the Enzymatic Mpemba effect to (B) the Markovian Mpemba effect. 
Each line is an energetic landscape of a single component. 
The amount of circles placed in each valley indicates the amount of components taking each state.
\label{fig:mpemba}
}
\end{figure}

The slowing down we have found is related to the Mpemba effect, which has recently received renewed attention \cite{Lasanta2017, Klich2019, Kumar2020, Carollo2021, Chatterjee2023, Rylands2024, Joshi2024, Shapira2024}, especially to the Markovian Mpemba effect \cite{Lu2017}. 
In the Markovian Mpemba effect, relaxation from an equilibrium state at one temperature to that at a lower temperature is faster when the initial temperature is higher. 
This counter-intuitive effect is explained by the imbalance at the equilibrium (Fig. \ref{fig:mpemba}B): 
Each component is obeyed by the multistable energy landscape and showes the stable, metastable, or intermediate state (in \ref{fig:mpemba}B, the stable, metastable, and intermediate states have an energy $E_1$, $E_2$, and $E_3$, respectively, where $E_1 < E_2 < E_3$).
This metastable state is surrounded by high energy barriers, and in order for the components in the metastable state to reach the stable state, it must cross a high energy barrier, which takes time. 
In the certain temperature range, when the temperature is lower, the probability of taking the metastable state at the equilibrium increases, and thus the relaxation to the equilibrium state at the lower temperature takes longer time.
This inbalance of the metastable state increases the distance to the final state from the equilibrium state at the lower temperature.

The slowing we found is explained by a similar mechanism  (Fig. \ref{fig:mpemba}A).
In a certain range of enzyme concentration, when the enzyme concentration is higher, the probability of entering the metastable state at the kinetic-memory state increases, and thus the relaxation to the equilibrium state with the higher enzyme concentration takes longer time. 
Because of the similarity, we named this phenomenon the Enzymatic Mpemba effect. 
The main difference between the Markovian Mpemba effect and the Enzymatic Mpemba effect is that the former is due to an imbalance in the equilibrium state, while the latter is due to an imbalance in the nonequilibrium state during slow relaxation. 
Our successful construction of the Enzymatic Mpemba effect based on the non-equilibrium state in the relaxation process indicates that non-intuitive relaxation phenomena such as the Mpemba effect are caused by various environmental variables, not just temperature. 
We expect that our discovery will be a stepping stone for the understanding of nontrivial relaxation phenomena in the future development of non-equilibrium physics.

\begin{acknowledgments}
The author would like to thank S. Ito for fruitful discussion and K. Kaneko for critical reading of the manuscript.
\end{acknowledgments}

\appendix*
\section{Derivation of the binding probability of an enzyme}
We consider the probability of binding of only a single enzyme.
Proteins with the different number of relaxed monomers have different binding energies, and a binding energy of the enzyme $j$ to a substrate in the $\sigma = i$ state is given by $E^{\mathrm{b}j}_i$.
Therefore, the grand partition function and the average number of binding enzymes $\langle N_{{\mathrm Enz}j} \rangle$ in equilibrium are represented in a similar manner as Langmuir's adsorption isotherm and are given by
\begin{eqnarray}
\Xi &=& \prod_{i=0}^M \left\{1 + \exp \left[ \beta \left(E^{\mathrm{b}j}_i + \mu\right) \right] \right\}^{N_i}, \\
\langle N_{\mathrm{Enz}j} \rangle &=& \sum_{i=0}^M N_i \frac{\exp(\beta \mu)}{\exp\left( -\beta E^{\mathrm{b}j}_i \right) + \exp(\beta \mu)}, \label{eq:avg_n}
\end{eqnarray}
where $N_i$ is the number of substrate in the $\sigma = i$ state.

For both {\it in vivo} and {\it in vitro} reactions, $N$ and $N_{\mathrm{Enz}j}$ are almost constant and are the control parameters of the system.
Here, we assume that $\langle N_{\mathrm{Enz}j} \rangle$ is the same as $N_{\mathrm{Enz}j}$, and that the chemical potential $\mu$ is a function of $N_{\mathrm{Enz}j}$ and $\{ N_i \}$.
This assumption is justified when $N$ and $N_{\mathrm{Enz}j}$ are sufficiently large, i.e., at the thermodynamic limit.
Then, the chemical potential $\mu$ can be calculated from Eq. \ref{eq:avg_n}.
Note that, in the limit $N_{\mathrm{Enz}j} \rightarrow N$, $\mu$ approaches $-\infty$, and all molecules bind to the enzyme.
The arguments so far can be summarized to give the binding probability
\begin{equation}
p^{\mathrm{b}j}_i = \frac{\langle N_{\mathrm{Enz}j, i} \rangle}{N_i} = \frac{\exp(\beta \mu)}{\exp\left( -\beta E^{\mathrm{b}j}_i \right) + \exp(\beta \mu)},
\end{equation}
where $N_{\mathrm{Enz}j, i}$ is the number of enzyme $j$ that bind to the a substrate in the $\sigma = i$ state.

\end{document}